\documentclass{elsart3p}
\usepackage{epsfig}
\newcommand{\be}{\begin{equation}}
\newcommand{\ee}{\end{equation}}
\newcommand{\bea}{\begin{eqnarray}}
\newcommand{\eea}{\end{eqnarray}}

\begin{document}

\begin{frontmatter}
\title{Quark matter nucleation in hot hadronic matter}
\author[Pisa]{I.~Bombaci},
\author[Pisa]{D.~Logoteta}, 
\author[Kolkata]{P.K.~Panda},
\author[Coimbra]{C.~Provid\^encia}
\author[Coimbra]{and I.~Vida\~na}

\address[Pisa]{Dipartimento di Fisica``E. Fermi'', Universit\`a di Pisa, and INFN, Sezione di Pisa, 
                      Largo B. Pontecorvo, 3
                      I-56127 Pisa, Italy }
\address[Kolkata]{Indian Association for the Cultivation of Sciences, 
 Jadavpur, Kolkata-700 032, India}
\address[Coimbra]{Centro de F\'{\i}sica Computacional, Department of Physics,
 University of Coimbra, 3004-516 Coimbra, Portugal}

\begin{abstract} 
We study the quark deconfinement phase transition in hot $\beta$-stable hadronic matter. 
Assuming a first order phase transition, we calculate the enthalpy per baryon of 
the hadron-quark phase transition. We calculate and compare the nucleation rate and the  nucleation time due to thermal  and quantum nucleation mechanisms. 
We compute the crossover temperature above which thermal nucleation dominates the finite temperature quantum nucleation mechanism. 
We next discuss the consequences for the physics of proto-neutron stars. We introduce the concept 
of limiting conversion temperature and critical mass  $M_{cr}$  for proto-hadronic stars,  
and we show that  proto-hadronic stars with a mass $M < M_{cr}$ could survive the early 
stages of their evolution without decaying to a quark star.   
\end{abstract}


\end{frontmatter} 

\section{Introduction}
In the last few years there has been a growing interest in the study of the nucleation 
process of quark matter (QM) in the core of massive neutron stars. 
In particular, it has been shown \cite{be03,bo04,drago04,lug05,blv07,bppv08,dps-b08,bambi08}  
that above a threshold value of the central pressure  a pure hadronic compact star (HS)  
is metastable to the decay (conversion) to a quark star (QS) ({\it i.e.} to a hybrid neutron star 
or to a strange star \cite{bod71,witt84}, depending on the details of the equation of state (EOS) 
for quark matter used to model the phase transition \cite{fj84,gle96,web05,haens_book}).    
This stellar conversion process liberates a huge amount of energy (a few $10^{53}$ erg) \cite{grb} 
and it could be the energy source of some of the long Gamma Ray Bursts (GRBs).       

The research reported in Refs \cite{be03,bo04,drago04,lug05,blv07,bppv08,dps-b08,bambi08}  
has focused on the quark deconfinement phase transition in cold ($T = 0$) and neutrino-free 
neutron stars. In this case the formation of the first drop of QM could take place solely via 
a quantum nucleation process. 

A neutron star at birth (proto-neutron star) is very hot (T =  10 -- 30 MeV) with neutrinos 
being still trapped in the stellar interior \cite{BurLat86,prak97}. 
Subsequent neutrino diffusion causes deleptonization 
and heats the stellar matter to an approximately uniform entropy per baryon $\tilde {S}$ =1 -- 2 
(in units of the Boltzmann's constant $k_B$). Depending on the stellar composition, during this stage 
neutrino escape can lead the more ``massive'' stellar configurations to the formation 
of a black hole \cite{bomb96,prak97}.  However, if the mass of the star is sufficiently small, 
the star will remain stable  and it will cool to temperatures well below 1~MeV  within 
a cooling time $t_{cool} \sim$~a~few~$10^2$~s, as the neutrinos continue to carry energy away 
from the stellar material \cite{BurLat86,prak97}.     
Thus in a proto-neutron star, the quark deconfinement phase transition will be likely triggered 
by a  thermal nucleation process. In fact, for sufficiently high temperatures, thermal nucleation 
is a much more efficient process with respect to the quantum nucleation mechanism.  

Some of the earlier studies on quark matter nucleation 
(see {\it e.g.}, \cite{ho92,ho94,ol94,hei95,harko04}) 
have already dealt with thermal nucleation in hot and dense hadronic matter. 
In these studies, it was found that the prompt formation of a critical size drop of quark matter 
via thermal activation is possible above a temperature of about $2-3$ MeV.  As a consequence, it was 
inferred that pure hadronic stars are converted to quark stars within the first seconds after their birth.  
However, these works \cite{ho92,ho94,ol94,hei95} reported an estimate of the thermal nucleation
based on "typical" values for the thermodynamic properties characterizing the central part of 
neutron stars.

Our main objective in this and next related works, is to establish if a newborn hadronic star 
(proto-hadronic star) could survive the early stages of its evolution without "decaying" to a quark star. 
In the present Letter, we calculate the thermal nucleation rate of quark matter in hot ($T \neq 0$) 
and neutrino-free hadronic matter using a finite temperature EOS for hadronic and quark matter.  
In addition, we calculate the quantum nucleation rate at finite temperature,   
and compare the thermal and quantum nucleation time at different temperatures and pressures  
characterizing the central conditions of metastable proto-hadronic compact stars.   
We compute the crossover temperature above which thermal nucleation dominates above the 
finite temperature quantum nucleation mechanism. 
Finally we briefly discuss some consequences for the physics of proto-neutron stars.

\section{Phase equilibrium}
 
For a first-order phase transition
{\footnote
{We assume the quark deconfinement phase transition to be of the first order. 
This assumption is common in most of the studies of quark deconfinement in compact stars.}}
the conditions for phase equilibrium  are given by the Gibbs' phase rule
\begin{equation}
T_H  = T_Q \equiv T   \, ,~~~~~~~~
P_H = P_Q \equiv P_0   
\label{eq:eq1a}
\end{equation}
\begin{equation}
\mu_H(T, P_0)  =  \mu_Q(T, P_0) \, 
\label{eq:eq1b}
\end{equation}

where 
\begin{equation}
  \mu_H = \frac{\varepsilon_H + P_H - s_H T}{n_H}  \, , ~
  \mu_Q = \frac{\varepsilon_Q + P_Q - s_Q T}{n_Q}  
\label{eq:eq2}
\end {equation} 
are the Gibbs' energies per baryon (average chemical potentials) for the hadron and 
quark phase respectively, 
$\varepsilon_H$ ($\varepsilon_Q$),  $P_H$ ($P_Q$), $s_H$ ($s_Q$)  and $n_{H}$  ($ n_{Q}$)
denote respectively the total ({\it i.e.,}  including leptonic contributions) energy 
density, total pressure, total entropy density,  and baryon number density  for the hadron (quark)  
phase.  
Above the "transition point" ($P_0$) the hadronic phase is metastable,  
and the stable quark phase will appear  as a results of a nucleation process. 

Small localized  fluctuations in the state variables of the metastable hadronic phase  
will give rise to virtual drops of the stable quark phase. These fluctuation are characterized 
by a time scale  $\nu_0^{-1} \sim 10^{-23}$ s. This time scale is set by the strong 
interactions (which are responsible for the deconfinement phase transition), 
and it is many orders of magnitude shorter than the typical time scale for the weak interactions.  
Therefore quark flavor must be conserved during the deconfinement transition.   
We will refer to this form of deconfined matter,  in which the flavor content is equal to that of 
the $\beta$-stable hadronic system at the same pressure and temperature, as the Q*-phase. 
Soon afterward a critical size drop of quark matter is formed, the weak interactions 
will have enough time to act, changing the quark flavor fraction of the deconfined droplet to lower 
its energy, and a droplet of $\beta$-stable quark matter is formed (hereafter the Q-phase).

This first seed of quark matter will trigger the conversion \cite{oli87,hbp91,grb} of the pure hadronic 
star to a hybrid star or to a strange star.  Thus, pure hadronic stars with values of the central pressure larger than  $P_0$ are metastable to the decay (conversion) to hybrid stars or to strange stars 
\cite{be03,bo04,drago04,lug05,blv07}. 
The mean lifetime of the metastable stellar configuration  is related to the time needed to nucleate the first drop of quark matter in the stellar center and  depends dramatically on the value of the 
stellar central pressure \cite{be03,bo04,drago04,lug05,blv07}.

\section{Quantum and thermal nucleation rates} 

The main effect of finite temperature on the quantum nucleation mechanism of quark matter 
is to modify the energy barrier separating the quark phase from the metastable hadronic phase.
This energy barrier, which represents the difference in the free energy of the system with and 
without a  Q*-matter droplet, can be written as 
\begin{equation}
  U({\cal R}, T) = \frac{4}{3}\pi n_{Q^*}(\mu_{Q^*} - \mu_H){\cal R}^3 + 4\pi \sigma {\cal R}^2
\label{eq:potential}
\end{equation}
where ${\cal R}$ is the radius of the droplet (supposed to be spherical), and $\sigma$ is 
the surface tension for the surface separating the hadron from the Q*-phase. 
The energy barrier has a maximum at the critical radius 
${\cal R}_c = 2 \sigma /[n_{Q^*}(\mu_H - \mu_{Q^*})]$.    
Notice that we have neglected the term associated with the curvature energy,   
and also the terms connected with the electrostatic energy, since they are known to introduce small corrections \cite{iida98,bo04}. 
The value of the surface tension $\sigma$ for the interface separating the quark and hadron phase 
is poorly known, and typically values used in the literature range within $10-50$ MeV fm$^{-2}$ 
\cite{hei93,iida97,iida98}.  We assume $\sigma$ to be temperature independent and we take 
$\sigma = 30$ MeV fm$^{-2}$.

The quantum nucleation time $\tau_q$ can be straightforwardly evaluated within a semi-classical 
approach \cite{lk72,iida97,iida98}. First one computes, in the WKB approximation, the ground state 
energy $E_0$ and the oscillation frequency $\nu_0$ of the drop in the potential well $U({\cal R},T)$.   
Then, the probability of tunneling is given by
\begin{equation}
  p_0=exp\left[-\frac{A(E_0)}{\hbar}\right]
\label{eq:prob}
\end{equation}
where $A(E)$ is the action under the potential barrier, which in a relativistic framework reads
\begin{equation}
 A(E)=\frac{2}{c}\int_{{\cal R}_-}^{{\cal R}_+}\sqrt{[2m({\cal R})c^2 +E-U({\cal R})][U({\cal R})-E]}   
\label{eq:action}
\end{equation} 
being ${\cal R}_\pm$ the classical turning points and $m({\cal R})$ the droplet effective mass. 
The quantum nucleation time is then equal to
\begin{equation}
  \tau_q  = (\nu_0 p_0 N_c)^{-1} \ , 
\label{eq:time}
\end{equation} 
with  $N_c \sim 10^{48}$  being  the number of nucleation centers expected in the innermost part 
($r \leq R_{nuc} \sim100$ m) of the hadronic star, where the pressure and temperature 
can be considered constant and equal to their central values.    

The thermal nucleation rate can be written \cite{LanTur73} as 
\begin{equation}
    I =\frac{\kappa}{2 \pi} \Omega_0 \exp (- U({\cal R}_c, T) /T)
\label{eq:therm_rate}
\end{equation}
where $\kappa$ is the so-called dynamical prefactor, which is related to the growth rate of the drop radius 
$\cal R$ near the critical radius (${\cal R}_c$),  $\Omega_0$ is the so-called statistical prefactor, which measures the phase-space volume of the saddle-point region around ${\cal R}_c$, and 
$U({\cal R}_c, T)$ is the activation energy, {\it i.e.} the change in the free energy of the system 
required to activate the formation of a critical size droplet. 
The Langer theory \cite{lang68,lang69,LanTur73,TurLan80} of homogeneous nucleation has been 
extended in Refs.\ \cite{CseKap92,VenVis94} to the case of first order phase transitions 
occurring in relativistic systems, as in the case of the quark deconfinement transition. 
The statistical prefactor, can be written \cite{CseKap92}  as 
\begin{equation}
   \Omega_0 = 
\frac{2}{3\sqrt{3}}  \Big(\frac{\sigma}{T}\Big)^{3/2}  \Big(\frac{\cal R}{\xi_Q}\Big)^4
\label{eq:omega}
\end{equation}
where $\xi_Q$ is the quark correlation length, which gives a measure of the thickness of the 
interface layer  between the two phases (the droplet "surface thickness"). 
In the present calculation we take  $\xi_Q = 0.7$~fm according to the estimate given in 
Refs.\ \cite{CseKap92,hei95}.   

For the dynamical  prefactor we have used a general expression which has been derived by 
Venugopalan and Vischer \cite{VenVis94} (see also Refs.\ \cite{CseKap92,CKO03})  
\begin{equation}
\kappa  = \frac{2 \sigma} {{\cal R}_c^3 (\Delta w)^2} \Big [ \lambda T + 2 \Big(\frac{4}{3} \eta + \zeta \Big)\Big]  \,  ,
\label{eq:kappa}
\end{equation}
where $\Delta w = w_{Q*} - w_H$ is the difference between the enthalpy density of the two phases, 
$\lambda$  the thermal conductivity, $\eta$ and $\zeta$ are the shear and bulk viscosities respectively 
of hadronic matter. Notice that the nucleation prefactor used in the present work differs significantly  
from the one used in previous works \cite{ho92,ho94,ol94}) where, based on dimensional grounds, the prefactor was taken to be equal to $T^4$.  

There are not many calculations of the transport properties of dense hadronic matter. 
With a few exceptions (see {\it e.g.} \cite{vDD04,CB06}),  
most of them are relative to nuclear or pure neutron matter \cite{FI79,Dan84,Sedr94,Benh07,Yakov08}.     
These quantities have been calculated by Danielewicz \cite{Dan84} in the case of nuclear matter.  
According to the results of Ref.\ \cite{Dan84}, the dominant contribution to the prefactor 
$\kappa$ comes from the shear viscosity $\eta$. Therefore, we take $\lambda$ and $\zeta$ equal to zero,
and we use for the shear viscosity the following relation \cite{Dan84}:   
\begin{equation}
\eta =  \frac{7.6 \times 10^{26}} {(T/{\rm MeV})^2}  \Big(\frac{n_H}{n_0}\Big)^2 ~~ 
            \frac{{\rm MeV}}{{\rm fm \, s}}  \,  ,
\label{eq:eta} 
\end{equation}                            
with $n_0 = 0.16$~fm$^{-3}$ being the saturation density of normal nuclear matter. 

 The thermal nucleation time $\tau_{th}$, relative to the innermost stellar region 
($V_{nuc} = (4 \pi/3) R_{nuc}^3$) where almost constant pressure and temperature occur, can thus 
be written as   $\tau_{th} = (V_{nuc} \, I )^{-1}$.

\section{Equation of state}

Over the last decade, it has been realized that strong interacting matter at high density  
and low temperature may possess a large assortment of  phases.  
Different  possible patterns for color superconductivity have been conjectured 
(see {\it e.g.} \cite{alf+08,CN04} and references therein quoted). 
Very recently, a new phase of QCD, named quarkyonic phase, has been predicted 
\cite{McP07,HMcP08}.  
This hypothetical matter phase is characterized by chiral symmetry and 
confinement \cite{McP07,HMcP08,McRS09}.

In the present work, we have adopted a more traditional view, assuming a single first order 
phase transition between the confined (hadronic) and deconfined phase of dense matter, 
and we have used  rather common models for describing them. 
For the hadronic phase we have used models which are based on a relativistic Lagrangian of hadrons interacting via the exchange of $\sigma$, $\rho$ and $\omega$ mesons. 
We have used one of the parameters sets  given in Ref.\ \cite{gm91}: hereafter we refer to this 
model as the GM1 equation of state.      
For the quark phase we have adopted a phenomenological EOS \cite{fj84} which is based on the MIT 
bag model for hadrons. In this work, we have used the following set of parameters: 
$m_u = m_d =0$, $m_s = 150$ MeV  for the masses of the {\it up}, {\it down } and {\it strange} quark 
respectively, $B = 85$ MeV/fm$^3$ for the bag constant,  and  $\alpha_s = 0$ for the 
QCD structure constant. 
The two models for the EOS have been generalized to the case of finite temperature.     

\begin{figure}[t]
\phantom{a}\vspace*{5mm}
{\psfig{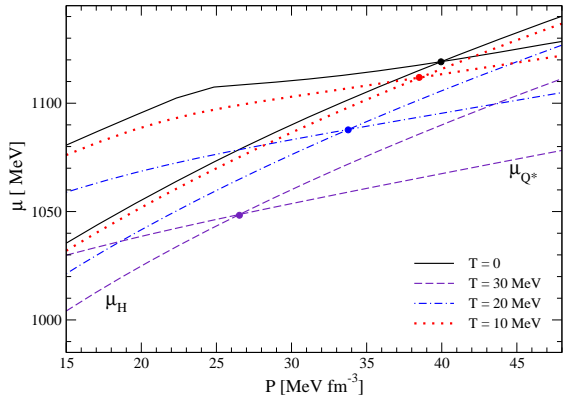}}
\caption{(Color online.) Gibbs energy per baryon as a function  of pressure for the hadronic and 
quark phases at different temperatures. Lines with the larger slope refer to the hadronic phase.
Full dots indicate the transition pressure $P_0$ for each temperature.} 
\label{Fig1}
\end{figure}

\section{Results and discussion}

In Fig. 1 we plot the Gibbs' energies per baryon for the hadron-phase and the Q*-phase at different temperatures, T = 0, 10, 20, 30 MeV;  lines with the larger slope refer to the hadron-phase.  
As we see, the transition pressure $P_0$ (indicated by a full dot) decreases when the hadronic 
matter temperature is increased. 
The phase equilibrium curve  $P_0(T)$ for the hadron-quark phase transition (within the present 
schematic model for the EOS) is shown in Fig. 2.  As it is well known, for a first-order phase transition 
the derivative $dP_0/dT$ is related to the specific  ({\it i.e.} per baryon) latent heat ${\cal Q}$  of the phase 
transition by the  Clapeyron-Clausius equation   
\begin{equation}
            \frac{dP_0}{dT} =  - \frac{ n_H n_{Q^*}}{n_{Q^*} - n_H } \frac{{\cal Q}}{T } 
\label{eq:eq10}
\end {equation}
\begin{equation}
           {\cal Q}  = \tilde W_{Q^*} - \tilde W_H = T(\tilde {S}_{Q^*} - \tilde {S}_H) 
\label{eq:eq11}
\end {equation}
where $\tilde W_H$ ($\tilde W_{Q^*}$)  and $\tilde {S}_H$ ($\tilde {S}_{Q^*}$) denote  respectively 
the  enthalpy  per baryon and entropy per baryon  for the hadron (quark) phase.   
The specific latent heat ${\cal Q}$ and the phase numbers densities  $n_H$ and  $n_{Q^*}$ 
at phase equilibrium are reported in Table 1.  As expected for a first order phase transition 
one has a discontinuity jump in the phase number densities: in our particular case 
$n_{Q^*}(T,P_0) > n_H(T,P_0)$. This result, together with the positive value of ${\cal Q}$ 
({\it i.e.} the deconfinement phase transition absorbs heat) tell us (see Eq.\ (\ref{eq:eq10})), 
that the transition temperature decreases with pressure (as in the melting of ice).  

\begin{figure}[t]
\phantom{a}\vspace*{5mm}
{\psfig{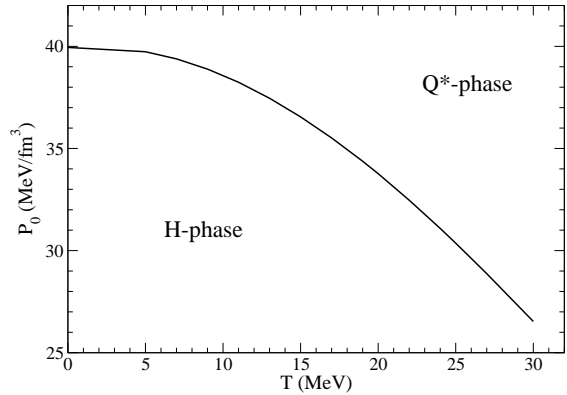}}
\caption{Phase equilibrium curve for the hadron to quark matter phase transition.} 
\label{Fig2}
\end{figure}
\begin{table}
\begin{tabular}{ccccc}
\hline
 $T$~~~ & ${\cal Q}$~~~~  & $n_{Q^*}$~~~~  & $n_{H}$~~~~ & $P_0$\\
 MeV~~~ & MeV~~~~ & fm$^{-3}$~~~~ & fm$^{-3}$~~~~ & MeV/fm$^3$ \\
\hline
  0~~~     &    0.00~~~~ &     0.453~~~~&  0.366~~~~  &     39.95 \\
  5~~~     &    0.56~~~~ &     0.451~~~~&  0.364~~~~  &     39.74 \\  
10~~~     &    2.40~~~~ &     0.447~~~~&  0.358~~~~  &     38.58 \\
15~~~     &    5.71~~~~ &     0.439~~~~&  0.348~~~~  &     36.55 \\
20~~~     &  10.60~~~~ &     0.428~~~~&  0.334~~~~  &     33.77 \\
25~~~     &  17.17~~~~ &     0.414~~~~&  0.316~~~~  &     30.36 \\
30~~~     &  25.44~~~~ &     0.398~~~~&  0.294~~~~  &    26.53 \\
\hline
\end{tabular}
\caption{The specific  latent heat ${\cal Q}$ and the phase numbers densities  $n_H$ and  $n_{Q^*}$ 
at phase equilibrium. } 
\label{table}
\end{table}

 In Fig. 3, we represent the energy barrier for a virtual drop of the Q*-phase  
as a function of the droplet radius and for different temperatures  at a fixed  pressure 
$P = 57$~MeV/fm$^3$.  As expected, from the results plotted in Fig. 1,  the energy 
barrier $U({\cal R}, T)$ and the droplet critical radius  ${\cal R}_c$ decrease as the matter 
temperature is increased. This effect favors the Q*-phase formation, and in particular increases 
(decreases) the quantum nucleation rate (nucleation time $\tau_q$) with respect to the 
corresponding quantities  calculated at  $T=0$.   

\begin{figure}[t]
\phantom{a}\vspace*{5mm}
{\psfig{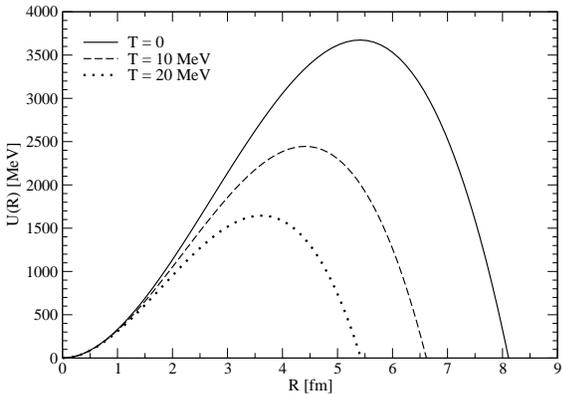}}
\caption{Energy barrier for a virtual drop of the Q*-phase  
as a function of the droplet radius and for different temperatures for a pressure 
$P = 57$~MeV/fm$^3$.} 
\label{Fig3}
\end{figure}

In Fig. 4  we plot the quantum and thermal nucleation times of the  Q*-phase  in  
$\beta$-stable hadronic matter as a function of temperature and at a fixed pressure $P=57$~MeV/fm$^3$. 
As expected, we find a crossover temperature $T_{co}$ above which thermal nucleation is dominant with respect to the quantum nucleation mechanism.  For the case reported in Fig. 4,  we have 
$T_{co}=7.05$~MeV and the corresponding nucleation time is $\log_{10}(\tau/{\rm s}) = 54.4$.
The crossover temperature, for different values of the pressure of $\beta$-stable hadronic matter,  
is reported in Tab. 2 (second column) together with the nucleation time calculated at  
$T = T_{co}$  (third column).  

\begin{figure}[t]
\phantom{a}\vspace*{5mm}
{\psfig{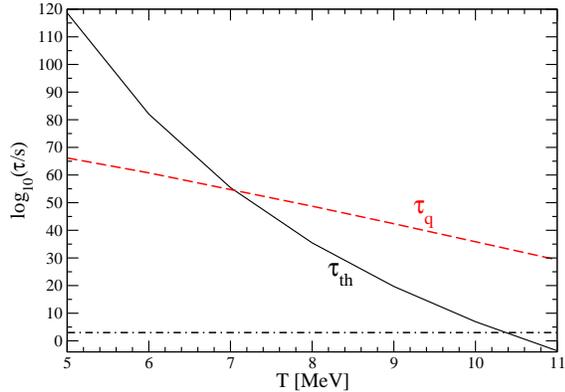}}
\caption{(Color online.) Thermal ($\tau_{th}$) and quantum ($\tau_q$) nucleation time of quark 
matter (Q*-phase) in $\beta$-stable hadronic matter as a function of temperature at fixed pressure 
$P= 57$~MeV/fm$^3$.  The crossover temperature is  $T_{co}=7.05$~MeV. 
The limiting conversion temperature for the proto-hadronic star is, in this case, $\Theta = 10.3$~MeV, 
obtained from the intersection of the thermal nucleation time curve (continuous line) and the 
dot-dashed line representing  $\log_{10}(\tau/{\rm s}) = 3$.}    
\label{Fig4}
\end{figure}
\begin{table}
\begin{tabular}{ccc}
\hline
$P$   & $T_{co}$ &~~$\log_{10}(\tau/{\rm s}) $  \\
\hline
   53.98~~~~&~~  5.0~~~~~&  233.6  \\
   55.48~~~~&~~  6.0~~~~~&  121.3  \\
   56.94~~~~&~~  7.0~~~~~&    56.6  \\
   58.42~~~~&~~  8.0~~~~~&    16.0  \\ 
   58.85~~~~&~~  8.3~~~~~&      3.0   \\
\hline
\end{tabular}
\caption{Crossover temperature $T_{co}$ (in MeV), for different fixed values of the 
pressure $P$ (in Mev/fm$^3$) of hadronic matter. The third column reports the logarithm 
of the nucleation time (in seconds) calculated at the crossover temperature. The 
value $8.3$ MeV defines the value of the limiting conversion temperature $\Theta$ for
a star with a central pressure $P=58.85$ MeV fm$^{-3}$.} 
\label{table}
\end{table}

Having in mind the physical conditions in the interior of a proto-hadronic star \cite{BurLat86,prak97} 
(see Sect. 1 of the present paper), to establish if this star will survive the early stages of its evolution 
without "decaying" to a quark star, one has to compare the quark matter nucleation time 
$\tau=\min(\tau_q,\tau_{th})$ with the cooling time  $t_{cool} \sim$~a~few~$10^2$~s.   
If $\tau >> t_{cool}$ then quark matter nucleation will not likely occur 
in the newly formed star, and this star will evolve to a cold deleptonized configuration.  
We thus introduce the concept of {\it limiting conversion temperature} $\Theta$ for the  
proto-hadronic star and define it as the value of the stellar central temperature $T_c$ for which 
the Q*-matter nucleation time is  equal to $10^3$~s. The limiting conversion temperature 
$\Theta$ will clearly depend on the value of the stellar central pressure (and thus on the value 
of the stellar mass).  

\begin{figure}[t]
\phantom{a}\vspace*{5mm}
{\psfig{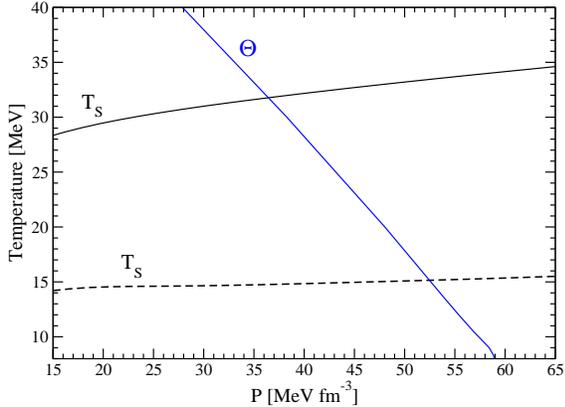}}
\caption{(Color online.) The limiting conversion temperature $\Theta$ for a newborn hadronic star 
as a function of the  central stellar pressure. 
Newborn hadronic stars with a central temperature and pressure located on the right side of 
the curve $\Theta(P)$ will nucleate a Q*-matter drop during the early stages of their evolution,  
and will finally evolve to cold and deleptonized quark stars, or will collapse to  black holes.  
The lines labeled  $T_S$ represent the stellar matter temperature 
as a function of pressure at fixed entropies per baryon $\tilde S/k_B = 1$ (dashed line) and $2$ (solid
line).}   
\label{Fig5}
\end{figure}

The limiting conversion temperature $\Theta$ is plotted in Fig. 5 as a function of the stellar 
central pressure.  A proto-hadronic star with a central temperature  $T_c > \Theta$  
will likely nucleate a Q*-matter drop during the early stages of its evolution,  
and will finally evolve to a cold and deleptonized quark star, or will collapse to a black hole 
(depending on the particular model adopted for the matter EOS).  

For an isoentropic stellar core \cite{BurLat86,prak97}, the central temperature of the 
proto-hadronic star is given, for the present EOS model, by the lines labeled by $T_S$ in Fig. 5,   
relative to the case $\tilde {S} = 1~k_B$ (dashed curve) and $\tilde {S} = 2~k_B$ (continuous curve).  
The intersection point ($P_S,\Theta_S$) between the two curves $\Theta(P)$ and $T_S(P)$
thus gives the central pressure and temperature of the configuration that we denote as 
the {\it critical mass} configuration of the proto-hadronic stellar sequence.  
The value of the gravitational critical mass $M_{cr} = M(P_S,\Theta_S)$  and baryonic critical mass 
$M_{B,cr}$ are reported in Tab.~3, for three different choices of the entropy per baryon,  
$\tilde {S}/k_B = 0$ (corresponding to a cold hadronic star)
{\footnote 
{Notice that in ref.\cite{be03,bo04,drago04,lug05,blv07,bppv08} the critical mass for cold ($T=0$) 
metastable hadronic stars has been defined as the value of the gravitational mass for 
which the quantum nucleation time is equal to one year:  $M_{cr}(T=0) = M(\tau_q=1~{\rm yr})$.   
It is worth recalling that the nucleation time $\tau_q$ is an extremely steep function of the 
hadronic star mass \cite{be03,bo04}, therefore the exact value of $\tau_q$ chosen in the definition of  
$M_{cr}(T=0)$ is not crucial (one must take a ``reasonable small'' value of $\tau_q$, much shorter 
that the age of young pulsars as the Crab pulsar). We have verified that changing  $\tau_q$ from 1~yr 
to $10^3$~s modifies $M_{cr}(T=0)$ by $\sim 0.02\%$.    
On the other hand, the nucleation time $\tau=\min(\tau_q, \tau_{th})$ entering in the definition 
of the critical mass of proto-hadronic stars $M_{cr}(\tilde{S})$ must be comparable to the  
proto-hadronic star cooling time $t_{cool}$. 
}}
, $1$ and $2$. 
In the same table, we also report the value of the gravitational mass ${\cal M}$ of the 
cold hadronic star with baryonic mass equal to $M_{B,cr}$.    
This configuration is stable ($\tau=\infty$) with respect to Q*-matter nucleation in the case   
$\tilde {S}/k_B = 2$, and it is essentially stable (having a nucleation time enormously larger 
than the age of the universe) in the case $\tilde {S}/k_B = 1$.  
Note that these numbers are model dependent and, therefore, one must take them just as indicative 
values. A careful analysis is beyond the scope of the present work and it will be addressed in a 
future work. 

\begin{table}
\begin{tabular}{cccc}
\hline
  $\tilde {S}/k_B$ & $M_{cr}$  & $M_{B,cr}$ & ${\cal M}$ \\
\hline
   0.0~~~~~  & 1.573~~~~~ & 1.752~~~~~ & 1.573 \\   
   1.0~~~~~  & 1.494~~~~~ & 1.643~~~~~ & 1.485 \\   
   2.0~~~~~  & 1.390~~~~~ & 1.492~~~~~ & 1.361 \\
\hline
\end{tabular}
\caption{Gravitational ($M_{cr}$) and baryonic ($M_{B,cr}$) critical mass (see text for more details) 
for proto-hadronic stars at different entropy per baryon  $\tilde {S}/k_B$.  
${\cal M}$ denotes the gravitational mass of  the cold hadronic configuration with the same 
stellar baryonic mass ($M_{B,cr}$).  
Stellar masses are in units of the solar mass, $M_\odot = 1.989 \times 10^{33}$~g.}   
\label{table}
\end{table}

In summary, in this work we have studied the quark deconfinement phase transition in hot 
$\beta$-stable hadronic matter, and we have explored some of its consequences for the physics 
of neutron stars at birth. Our main finding is that proto-hadronic stars with a mass lower that 
the critical value $M_{cr}$ could survive the early stages of their evolution without 
decaying to a quark star.  
However, the prompt formation of a critical size drop of quark matter could take place when 
$M > M_{cr}$.  These proto-hadronic stars evolve to cold and deleptonized quark stars, 
or collapse to a black holes. Finally,  if quark matter nucleation occurs during post-bounce stage 
of core-collapse supernova, then the quark deconfinement phase transition could trigger a 
delayed supernova explosion characterised by a peculiar neutrino signal \cite{sage+09}.

\section*{Acknowledgement}
This work was partially supported by FCT (Portugal) under the project  CERN/FP/83505/2008 
and by COMPSTAR, an ESF Research Networking Programme.   


\end{document}